\documentclass[aps,prl,twocolumn,nopacs,floatfix]{revtex4-2}
\usepackage[utf8]{inputenc}
\usepackage{textcomp}
\usepackage{epsfig}
\usepackage{graphicx} 
\usepackage{dcolumn}
\usepackage{amsthm,amsmath}
\usepackage{comment}
\usepackage[colorlinks=true, allcolors=blue]{hyperref}
\usepackage{xcolor}
\usepackage{colortbl}
\usepackage{multirow}
\usepackage{bigdelim}
\usepackage{colortbl}
\usepackage{mathtools}
\usepackage{orcidlink}
\usepackage{amsmath}
\usepackage[normalem]{ulem}
\usepackage{bm}
\usepackage[T1]{fontenc}
\usepackage{orcidlink}
\usepackage{graphicx}% Include figure files
\usepackage{dcolumn}% Align table columns on decimal point
\usepackage{bm}% bold math

\begin{document}

\preprint{APS/123-QED}

%\title{Revisiting King-plot Nonlinearities in the Singly-ionized Calcium using Improved Second-order Isotope Shift Constants}
%\title{Reappraising Constraints on the Mass of a Vector Boson from the King's Plot by Rigorously Evaluating the Second-order Isotope Shift Constants in Ca$^+$}

\title{Comprehensive Inclusion of Higher-order Ca$^+$ Isotope Shifts in the King's Plot Yields an Order Improvement on the $e^- - n$ Coupling Limit}

%\title{Analysis of Higher-order Ca$^+$ Isotope Shifts for improved e$^-$- n coupling bound}

\author{$^{a,b}$Vaibhav Katyal\orcidlink{0000-0002-7717-8558}}
\email{vaibhavkatyal@prl.res.in}
\author{$^{a}$A. Chakraborty\orcidlink{0000-0001-6255-4584}}
\author{$^a$B. K. Sahoo\orcidlink{0000-0003-4397-7965}}
\email{bijaya@prl.res.in}

\affiliation{
$^a$Atomic, Molecular and Optical Physics Division, Physical Research Laboratory, Navrangpura, Ahmedabad 380009, India}  
\affiliation{
$^b$Indian Institute of Technology Gandhinagar, Palaj, Gandhinagar 382355, India
}

\begin{abstract}
By critically evaluating higher-order nonlinear effects to the isotope shifts (ISs) in the low-lying transition frequencies of the singly charged calcium ion, stringent constraint on the electron-neutron coupling due to a hypothetical boson describing physics beyond the Standard Model is inferred. It shows an order magnitude difference compared to the previously reported limit demonstrating importance of higher-order effects in the analysis of nonlinearity in the King's plot. The first-order IS parameters and enhancement factor ($D$) were evaluated using two complementary approaches in the relativistic coupled-cluster theory framework: namely finite-field (FF) and analytical response (AR) approaches. Extraction of the second-order IS parameters in the FF approach show numerical instabilities, so they are determined in the AR approach. Comparison of these factors with previous calculation shows substantial differences in the magnitudes. However, $D$  values from both the FF and AR approaches display excellent agreement. We also show explicitly roles of electron correlation effects in the evaluation of $D$ values accurately. 
\end{abstract}

\maketitle

Extensive searches for new physics (NP) beyond the Standard Model (SM) have been pursued in different areas, ranging from particle physics to astrophysics, with particular interest in search of a new boson that could serve as a dark matter candidate \cite{raffelt2012limits,choi2021recent, essig2009probing,  demille2017probing}. Traditionally, much of this effort has focused on weakly interacting massive particles with masses $\gtrsim 10$ GeV/${\rm c^2}$. More recently, however, growing attention has turned to an alternative class of models that postulate existence of a lighter boson ($\phi$), with mass $\lesssim 100$ MeV/${\rm c^2}$, mediating interactions between electrons and neutrons \cite{delaunay2017probing, berengut2018probing}. A distinctive signature of such a boson can be found in isotope shift (IS) measurements. Considering IS is dominated by the first-order mass-shift (MS) and field-shift (FS), which can be factorized into isotope-dependent and electronic-transition-dependent terms; we can get a linear equation of the differential ISs in two transitions which is referred to as the King's plot equation \cite{king2013isotope}. However, the exchange of a new boson, which represents physics beyond the SM (BSM), can cause nonlinearity (NL) in the King's plot \cite{delaunay2017probing}. To extract NL effect from the King's plot due to this BSM particle, it is equally important to quantify other NL effects arising from the SM.   

The linear King's plot equation for IS between two transitions, denoted by $a$ and $b$, in two isotopes $A$ and $A'$, is given by \cite{king2013isotope}
\begin{equation}\label{linear}
\delta \nu^{AA'}_{a} \approx \left(K_{a}-K_{b}\frac{F_{a}}{F_{b}}\right) \mu^{AA'} +\frac{F_{a}}{F_{b}}\delta\nu^{AA'}_{b},
\end{equation}
where $\delta\nu^{AA'}_{a}$ and $\delta\nu^{AA'}_{b}$ are the ISs of $a$ and $b$ transitions, respectively, $\mu^{AA'}=\frac{1}{M_{A'}}-\frac{1}{M_A}$ with $M_{A/A'}$ being the mass of the isotopes $A/A'$, $K$s and $F$s denote MS and FS constants, respectively. The MS constant is divided into normal MS (NMS) and specific MS (SMS) constants; i.e. $K=K_{NMS}+K_{SMS}$. Definitions of the $F$, $K_{NMS}$ and $K_{SMS}$ operators can be found in Ref. \cite{sahoo2025recent}. 

Presence of electron-neutron ($e^- -n$) interaction by the BSM vector boson can produce an additional shift $(A-A') D_{ab} \alpha_{NP}$ in the above expression with $\alpha_{NP}=\frac{y_ey_n}{4\pi c}$ is the $e^- -n$ coupling for the electron and neutron coupling strengths $y_e$ and $y_n$, respectively, and $D$ is defined as (in atomic units (a.u.))
\begin{equation}
D=\left(-1\right)^{s+1}\int\text{d}^3r~\rho_e(r)\frac{e^{-m_{\phi}r/c}}{r} ,
\end{equation}
where $s$ and $m_{\phi}$ are spin and mass of the boson, respectively, $c$ denotes speed of light and $\rho_e(r)$ is the electron density function normalized to unity.

Calcium (Ca) has five long-lived even mass number ($A=40$, 42, 44, 46, 48) isotopes and their singly charged ions (Ca$^+$) have very simple atomic structure. This enables to carry out high-precision IS measurements in the low-lying transitions of Ca$^+$ ions. Absence of hyperfine structures owing to their zero-nuclear spin also make solid basis for carrying out King's plot analysis unambiguously in these isotopes. Recent IS measurements in these ions have shown intriguing outcome in which King's plots still look linear even when uncertainties reach 10 Hz level \cite{solaro2020improved, chang2024systematic, knollmann2023part}. This persistence implies NL effects remain below current sensitivity, although deviations from linearity may become apparent as IS measurement precision advances towards the 1 Hz regime. It is, therefore, imperative to estimate these SM NL contributions to IS scrupulously to search for any plausible NP \cite{flambaum2018isotope}. To do this, a more general King's plot equation is defined by \cite{delaunay2017probing, hur2022evidence}
\begin{eqnarray}\label{nonlinear_IS}
\delta \nu_{a}^{AA'} & \simeq & \left [ K_{ab} \mu^{AA'} + F_{ab} \delta \nu_{b}^{AA'} \right ]  + G_{ab}^{(2)} \left(\delta\langle r^{2} \rangle^2 \right)^{AA'} \nonumber \\ && + G_{ab}^{(4)} \delta \langle r^{4}\rangle_{AA'} + \ F_{ab}^{(2)} \left(\delta\langle r^{2} \rangle^2 \right)^{AA'} \nonumber \\
&& + K^{(2)}_{ab} \Tilde{\mu}^{AA'} + (A-A') D_{ab} \alpha_{NP} ,
\end{eqnarray}

\begin{table*}[t!]
\setlength{\tabcolsep}{1pt}
\caption{The first-order NMS, SMS and FS constants of the low-lying transitions in Ca$^+$ using our RCCSDT method and other corrections in both the AR and FF approaches. These values are compared with the literature values. We have assigned uncertainties to our recommended FF results. Value marked with `*' is the experimental result from Ref. \cite{gebert2015precision}.}
\begin{ruledtabular}\label{IS_FF}
\begin{tabular}{l ccc ccc ccc}
Transition & \multicolumn{3}{c}{$K_{NMS}$ (GHz amu)}  & \multicolumn{3}{c}{$K_{SMS}$ (GHz amu)}  & \multicolumn{3}{c}{$F$ (MHz/fm$^{2}$)}  \\[1ex]
\cline{2-4} \cline{5-7} \cline{8-10} \\
&  AR  & FF & Scaling \cite{NIST}  & \multicolumn{1}{c}{AR} & \multicolumn{1}{c}{FF} & Others  & \multicolumn{1}{c}{AR} & \multicolumn{1}{c}{FF} & Others \\ 
\hline \\
$4S - 3D_{3/2}$ & 171 & 227(3) & $227$ & 2221 & 2220(5) & 2236 \cite{viatkina2023calculation}, $2342$ \cite{safronova2001third} &  $-369$ &  $-370(3)$ & $-379$ \cite{viatkina2023calculation}, $-378$ \cite{safronova2001third} \\
%& & & & & & $2342^b$\\ [1ex]
$4S - 3D_{5/2}$  & 175 &  $223(3)$  & $225$ & 2212 & $2215(5)$ & $2223$ \cite{viatkina2023calculation}, $2336$ \cite{safronova2001third}   & $-368$ & $-365(3)$  & $-378$ \cite{viatkina2023calculation} \\ 
%& & & & & & $2336^b$\\ [1ex]
$4S - 4P_{1/2}$ & 410 & $413(2)$ & $414$ & $-28$ & $-20(5)$  & $-30$ \cite{viatkina2023calculation}, $-55$ \cite{safronova2001third} & $-280$ & $-280(1)$ & $-281.8(7)^*$ \cite{gebert2015precision} \\ 
%&  &  &  &  & & $-55^b$ & &  $-286.4$ \cite{xyz} \\[1ex]
$4S - 4P_{3/2}$ & 411  & $415(2)$ & $418$ & $-32$ & $-23(5)$  & $-45$ \cite{viatkina2023calculation}, $-59$ \cite{safronova2001third} & $-280$ & $-280(1)$ & $-286.7$ \cite{viatkina2023calculation} \\
%& & & & & & $-59^b$\\ [1ex]
$3D_{3/2} - 4P_{1/2}$ & 240 & 186(3) & 190 &  $-2249$ & $-2239(5)$  & $-2266$ \cite{viatkina2023calculation},  $-2397$ \cite{safronova2001third} & $89$ & $90(1)$ & 92.6 \cite{viatkina2023calculation}, 92 \cite{safronova2001third}\\
%& & & & & & $-2397^b$\\ [1ex]
%$3D_{3/2} - 4P_{3/2}$ &  241 & 188 & 194 & $-2253$ & $-2242$ & $-2281^a$\\
%& & & & & & $-2401^b$\\ [1ex]
%$3D_{5/2} - 4P_{1/2}$ &  235 & 190 & 189 & & $-2234$ & $-2253^a$\\
%& & & & & & $-2391^b$\\ [1ex]
%$3D_{5/2} - 4P_{3/2}$ &  237 & 192 & 193 & & $-2237$ & $-2268^a$\\
%& & & & & & $-2395^b$\\ [1ex]
\end{tabular}
\end{ruledtabular}
%Ref: $^a$\cite{viatkina2023calculation}, $^b$\cite{safronova2001third}
\end{table*}
where $\Tilde{\mu}^{AA'}=\frac{1}{M_A^2}-\frac{1}{M_{A'}^2}$, $G^{(2)}$ are the second-order FS constants due to second Barrett radii, $G^{(4)}$ are another first-order FS constants due to fourth Barrett radii, $F^{(2)}$ are the second-order FS constants arising from the perturbed wave functions due to the FS operator, $K^{(2)}$ are the second-order MS constants arising from the perturbed wave functions due to the MS operator. $X_{ab}$ for any quantity $X$ is defined as $X_{ab}= X_a-\frac{F_a}{F_b} X_b$ except for the definition $F_{ab}=\frac{F_a}{F_b}$. Analogous to the first-order MS constant, $K^{(2)}$ is also determined separately as the second-order NMS and SMS constants; i.e. $K^{(2)}=K_{NMS}^{(2)}+K_{SMS}^{(2)}$. We adopt the operator form of $G^{(2)}$ and the analytical form of $G^{(4)}$ from Ref. \cite{hur2022evidence} for the analysis. The IS arising from  $G^{(4)}$ depends on the coefficient of $r^2$ in the power series expansion of $\rho_e(r)$. To obtain this coefficient, we compute $\rho_e(r)$ using the relativistic coupled-cluster (RCC) theory and fit it to a power series for extracting the $r^2$ term \cite{hur2022evidence}. 

For the determination of the $F$, $K$, $G^{(2)}$, $G^{(4)}$ and $D$ constants, very accurate wave functions, due to atomic Hamiltonian ($H_{at}$) encapsulating all possible SM interactions, are required. However, first-order perturbed wave functions due to respective IS operators ($O^{IS}$) are also needed to estimate the $F^{(2)}$ and $K^{(2)}$ constants. Gebert {\it et al} \cite{gebert2015precision} experimentally determined $F$ and $K^{MS}$ for the $4S \rightarrow 4P_{1/2}$ and $3D_{3/2} \rightarrow 4P_{1/2}$ transitions. The most recent calculations on both the first- and second-order IS constants in Ca$^+$ ion were performed by Viatkina {\it et al.} \cite{viatkina2023calculation}, but $D$ values for this ion were not given explicitly earlier. In fact, uncertainties of the calculated $F$ and $K^{MS}$ constants of Ca$^+$ are not yet ascertained. In a recent work \cite{chakraborty2025investigating}, we have reported very accurate values of energies, hyperfine structure constants, lifetimes and polarizabilities of the clock states of Ca$^+$ by employing RCC theory, assumed to be Gold standard of many-body methods, at the singles, doubles and triples excitation approximation (RCCSDT method). Here, we first attempt to improve accuracies of the $F$, $K^{NMS}$, $K^{SMS}$ and $D$ constants using the RCCSDT method, then extend this method to evaluate the $G^{(2)}$, $G^{(4)}$, $F^{(2)}$ and $K^{(2)}$ constants of the $4S$, $3D_{3/2,5/2}$ and $4P_{1/2,3/2}$ states of Ca$^+$. 

\begin{figure}[b]
\centering
\begin{tabular}{cc}
\includegraphics[width=42mm,height=47mm]{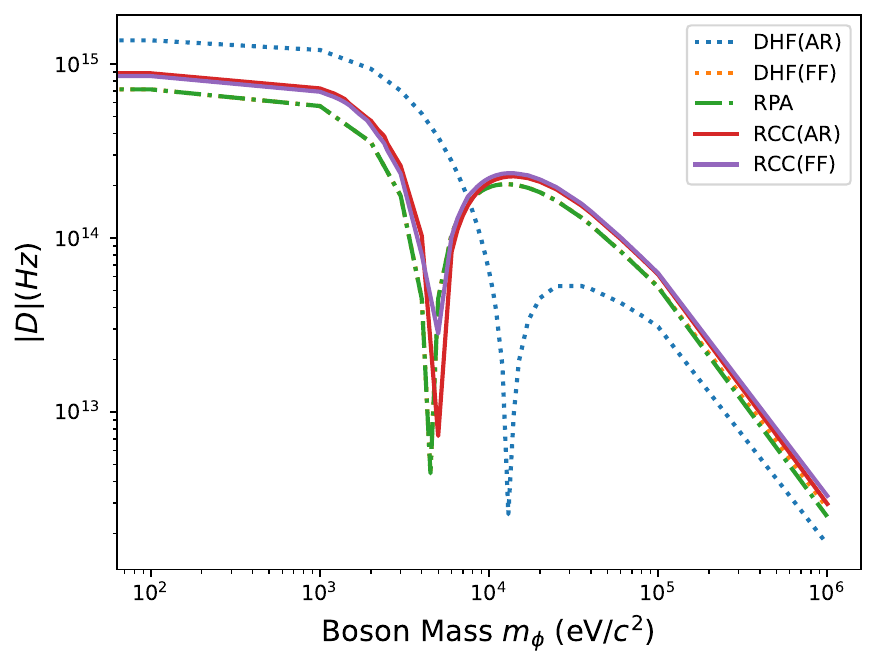}  &  \includegraphics[width=42mm,height=47mm]{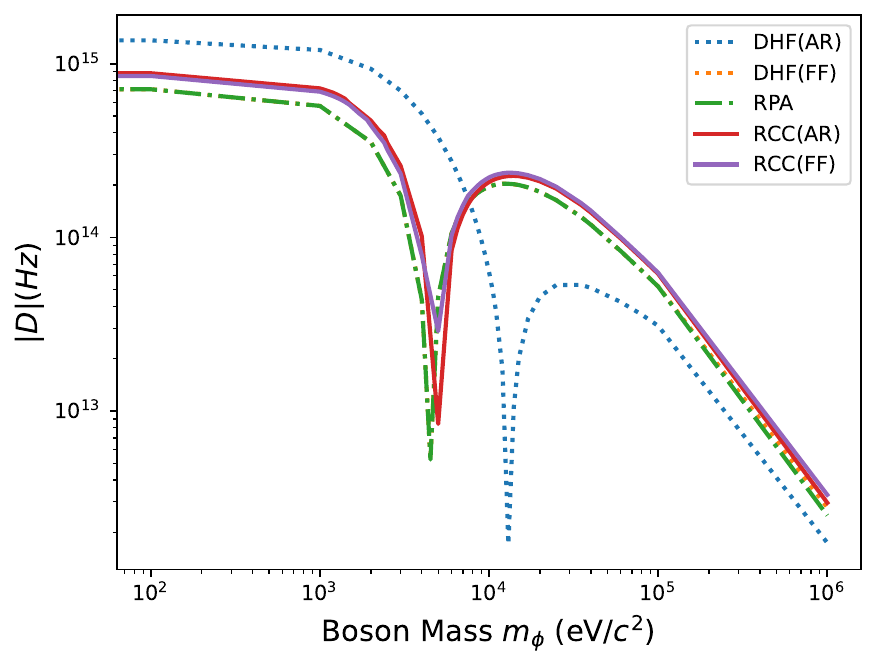}\\
 (a) & (b)\\
 \end{tabular}
\caption{Plot demonstrating $D$ values (in Hz) for the (a) $4S\rightarrow 3D_{3/2}$ and (b) $4S_{1/2}\rightarrow3D_{5/2}$ transitions in Ca$^+$ with respect to $m_{\phi}$ (in eV/c$^2$) from the DHF, RPA and RCC methods of the AR and FF approaches.}
\label{Dvsm}
\end{figure}

\begin{table*}[t!]
\caption{Higher-order FS and MS constants for different transitions in the Ca$^+$ ion using the DHF and RCC methods. Comparison of these values from the RCC method with those reported in Ref. \cite{viatkina2023calculation} show an order of magnitude difference.}
\begin{ruledtabular}\label{CC-SMNL}
\begin{tabular}{l ccc cc ccc cc ccc}
Transition & \multicolumn{3}{c}{$G^{(2)}$} & \multicolumn{2}{c}{$G^{(4)}$ } &  \multicolumn{3}{c}{$F^{(2)}$}  & \multicolumn{2}{c}{$K^{(2)}_{NMS}$} & \multicolumn{3}{c}{$K^{(2)}_{SMS}$} \\
& \multicolumn{3}{c}{(in MHz/fm$^4$)} & \multicolumn{2}{c}{(in KHz/fm$^4$)} & \multicolumn{3}{c}{(in KHz/fm$^4$)}  & \multicolumn{2}{c}{ (in GHz-amu$^2$)} & \multicolumn{3}{c}{(in GHz-amu$^2$)} \\
\cline{2-4} \cline{5-6} \cline{7-9} \cline{10-11} \cline{12-14} \\
 & DHF & RCC & Ref.~\cite{viatkina2023calculation} & DHF & RCC & DHF & RCC & Ref.~\cite{viatkina2023calculation} & DHF & RCC & DHF & RCC & Ref.~\cite{viatkina2023calculation}  \\
\hline \\
$4S\;-\;3D_{3/2}$ & 2.084 & $3.55(20)$ & $0.07$ & $-10.15$ & $-17.2(7)$ & $-0.17$ & $-0.22(2) $ & $-126.75$ & 26.37 & $-3.5(4) $ & 18.68 & $29(2)$ & $1.96$ \\
$4S\;-\;3D_{5/2}$ & 2.084 & $3.54(20)$ & $0.07$ & $-10.15$ & $-17.2(7)$ & $-0.17$ & $-0.22(2)$ & $-126.55$ & 26.23 &  $-3.5(4)$ & 18.68 & $28(2)$  & $1.93$\\
$4S\;-\;4P_{1/2}$ & 2.079 & 2.72(18) & 0.05 & $-10.13$ & $-13.3(5)$ & $-0.17$ & $-0.25(3)$ & $-95.83$ & $-2.74$ & $-1.58(7)$ & 0.58 & $7.6(5)$ & $5.72$\\
$4S\;-\;4P_{3/2}$ & 2.084 & $2.72(18)$ & $0.05$ & $-10.15$ &  $-13.3(5)$ & $-0.17$ & $-0.25(3)$ & $-95.9$ & $-2.77$ & $-1.62(7)$ & 0.61 & $7.7(5)$ & $5.73$ \\
$3D_{3/2}\;-\;4P_{1/2}$ & $\sim 0$ & $-0.83(7)$ & $-0.01$ & $\sim 0$ & 3.96(12) & $\sim 0$ & $-0.03(0)$ & 30.92 & $-29.11$ & $1.89(8)$ & $-18.11$ & $-22(2)$  & $3.78$\\
\end{tabular}
\end{ruledtabular}
\end{table*}

To ensure reliability in the calculations of IS constants, we consider both the finite-field (FF) and analytical response (AR) approaches in the RCC theory framework which are complementary to each other \cite{sahoo2025recent}. In the FF approach, the first- and second-order IS constants are estimated following the Taylor's expansion
\begin{eqnarray}
E_v(\lambda) = E_v(0) + \lambda \langle O^{IS} \rangle+ \frac{\lambda^2}{2} \langle O^{IS} \rangle^{(2)}  + {\cal O}(\lambda^3)  ,
\label{eqn1}
\end{eqnarray}
where $E_v(0)$ is the energy of an atomic state with valence orbital $v$ due to $H_{at}$, $E_v(\lambda)$ is the modified energy due to the total atomic Hamiltonian $H_{\lambda} = H_{at} + \lambda O^{IS}$. {\it Albeit} comparison of the calculated $E_v(0)$ with measurements may look very good, it does not ensure accuracy in the extracted $\langle O^{IS} \rangle$ and $\langle O^{IS} \rangle^{(2)}$ constants from the fitting of the $E_v(\lambda)$ values for different $\lambda$s in Eq. (\ref{eqn1}). This is owing to the fact that numerical differentiations in the FF approach are very sensitive to choice of $\lambda$ for different constants and states. To circumvent this problem, we determine the first-order IS constant and wave function due to $O^{IS}$ for a state $|\Psi_v^{(0)} \rangle$ by solving the inhomogeneous differential equation in the AR approach \cite{sahoo2025recent,Katyal2025, chakraborty2025investigating}
\begin{eqnarray}
(H_{at}-E_v^{(0)} |\Psi_v^{(1)} \rangle = (\langle O^{IS} \rangle - O^{IS} ) |\Psi_v^{(0)} \rangle ,
\label{eqn2}
\end{eqnarray}
where $|\Psi_v^{(1)} \rangle$ is its first-order perturbed wave function due to the respective IS operator $O^{IS}$. The $\lambda$ independent expression in the AR approach is a better choice for evaluating the IS constants, but accuracy of its results are limited by how well electron correlation effects are accounted through the RCCSDT method. In the FF approach, core-polarization effects (also known as orbital relaxation effects) are included to all-orders at the starting level of Dirac-Hartree-Fock (DHF) calculation ensuing to account for more correlations at the RCCSDT method. Consequently, the IS constants obtained using the RCCSDT method in the FF approach can be safely considered to be more accurate than the results obtained from the AR approach provided numerical uncertainties due to choice of $\lambda$ in the FF are negligibly small. To ensure this, we compare the RCCSDT values from both the approaches for all the states and properties.  

In Table~\ref{IS_FF}, we present {\it ab initio} results for the differential $K$ and $F$ constants for five low-lying transitions of Ca$^+$ using both the AR and FF approaches in the RCCSDT method, and compare them with the earlier reported values and those are inferred from the IS measurements using the King's plot \cite{safronova2001third, wilzewski2412nonlinear, gebert2015precision, berengut2003isotope, viatkina2023calculation}. We have considered the Dirac-Coulomb-Breit interactions and lower-order vacuum polarization effects in $H_{at}$ for the calculations as discussed in Ref. \cite{sahoo2025recent}. Uncertainties are estimated only to the FF values for their later use. We find excellent agreement between results from both the AR and FF approaches of this work, except for $K^{NMS}$ where orbital relaxation effects are found to be very strong. We also observe reasonable agreement with the previous calculations \cite{viatkina2023calculation, safronova2001third, maartensson1992isotope}.

We plot the $D$ values against $m_{\phi}$ from the DHF and RCC methods of the AR and FF approaches in Fig. \ref{Dvsm}, which shows substantial differences between the DHF values from both the approaches. Since the DHF values in the FF approach embeds orbital-relation effects, we include them through random-phase approximation (RPA) to verify the DHF values of the AR approach. As can be seen in the above figure, the RPA results and DHF values of the FF approach overlap suggesting that our implementation of both the approaches are correct. From the differences between the RPA and RCC calculations in the above plots, it is evident that non-RPA correlation contributions to $D$ are small though non-negligible. 

The calculated $G^{(2)}$, $G^{(4)}$, $F^{(2)}$, $K_{NMS}^{(2)}$ and $K_{SMS}^{(2)}$ values from this work are listed in Table \ref{CC-SMNL}. Since these constants correspond to extremely tiny contributions to ISs, we observe critical numerical errors in estimating them in the FF approach. Therefore, these higher-order contributing IS constants are evaluated using the AR approach, except for the $G^{(4)}$ values where we adopted the earlier mentioned approach. Uncertainties to these values are gauged from the first-order IS constant calculations. Another calculation reported in Ref. \cite{viatkina2023calculation}, provides values only for $G^{(2)}$, $F^{(2)}$ and $K_{SMS}^{(2)}$. In their work $G^{(2)}$ is defined to be proportional to the FS constant $F$, with the proportionality factor $f_{ho}$ that remains same for all five valence states of Ca$^+$, but varies for different isotope pairs. For the $A=40$ and $A'=46$ isotope pair, our calculation yields $f_{ho}=1.17$ in contrast to $f_{ho}=0.0238$ in Ref. \cite{viatkina2023calculation}. It appears to us that a variant of the FF approach was employed in Ref. \cite{viatkina2023calculation} to determine these constants, resulting in values approximately an order of magnitude smaller than ours. We could not decipher the exact reason for such large differences in both the calculations, but we attribute this to numerical problem of the FF approach for estimating the second-order effects. 

\begin{table}[t!]
\caption{The extracted constraint on the coupling constant $|\alpha_{NP}| \times 10^{-10}$ at $2\sigma$ level for a few selected $m_\phi$ values using both the AR and FF approaches in the Case I and Case II analyses. Our Case I data agree reasonably with the earlier analyses for smaller value of $m_{\phi}$, but there are significant differences in the values when Case II is considered.}
\begin{ruledtabular}\label{NP_constraint}
\begin{tabular}{l cc ccc}
    $m_{\phi}$ & \multicolumn{2}{c}{Case I} & \multicolumn{2}{c}{Case II} & \multicolumn{1}{c}{Case I} \\
\cline{2-3}  \cline{4-5}\\
(eV) &  AR & FF  &  AR  & FF & Others (FF) \\ 
\hline \\
0   &  0.79  & 0.80  & 3.9 &  5.9 &  $0.69$ \cite{solaro2020improved}, 0.14 \cite{chang2024systematic}  \\
%     &   &   &  & &  0.14 \cite{chang2024systematic} \\
100  & 0.79 & 0.78 & 3.9  & 5.9  &  \\
1000  & 1.01 & 1.1 &  4.7 & 7.3 & \\
$10^4$ & 2.19 & 1.9 & 17 & 23  &  \\
$10^5$ & 9.1 &  11 & 56 & 81  & \\
$10^6$ & 198 &  173  & 1164 & 1534  & \\
\end{tabular}
\end{ruledtabular}
\end{table}

\begin{figure}[t!]
\centering
\includegraphics[width=\linewidth]{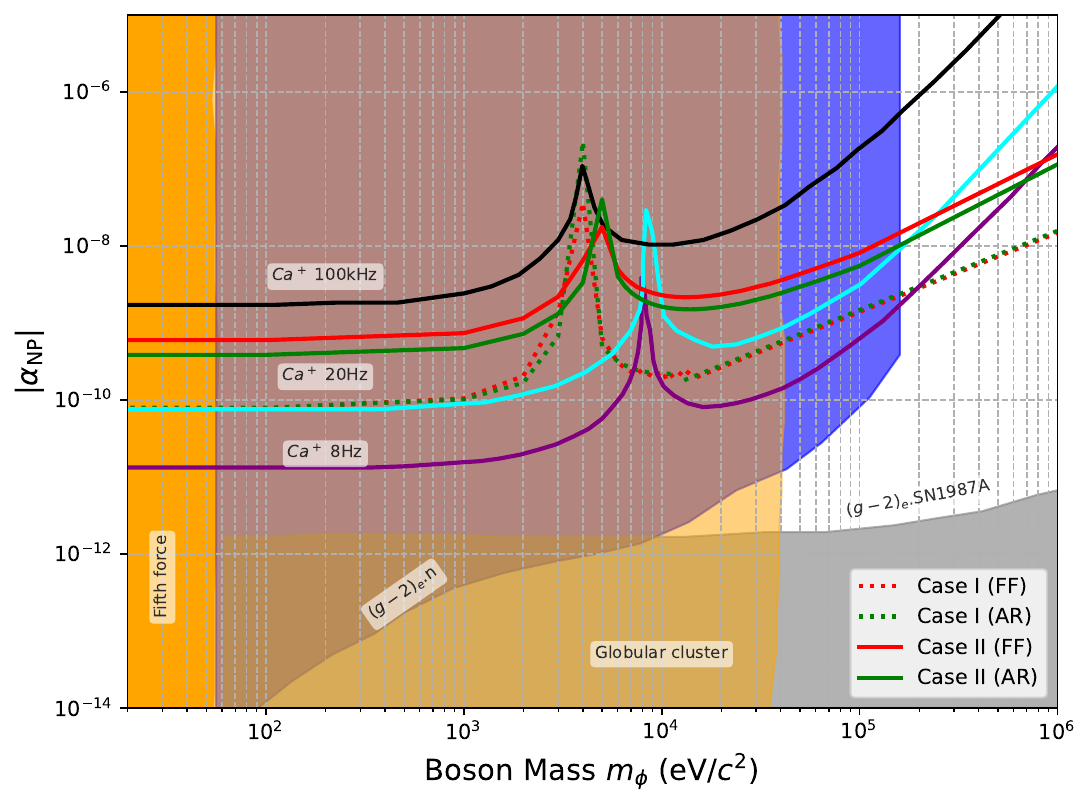}
\caption{Exclusion plot demonstrating constraints (2$\sigma$ upper bound) on $|\alpha_{NP}|$ with respect to $m_{\phi}$. The red and green solid line represent limits imposed using the FF results through Case I and Case II analyses from the present work, respectively, black \cite{gebert2015precision}, cyan \cite{solaro2020improved} and purple \cite{chang2024systematic} solid lines are the constraints from the previous IS studies, while constraints from other complementary studies such as Casimir effects \cite{bordag2009advances}, $(g-2)$ factor of the electron \cite{fan2023measurement, morel2020determination} and neutron scattering \cite{nesvizhevsky2008neutron} and star cooling in a supernova \cite{raffelt2012limits} are also shown.}
\label{bound}
\end{figure}

Since Eq.~(\ref{nonlinear_IS}) contains five independent sources of nonlinearity, at least seven transitions in eight independent isotope pairs would be required to constrain $\alpha_{NP}$ \cite{hur2022evidence}. It is intricate to achieve this target as measuring ISs on equal precision level for so many transitions is insurmountable. In the previous analyses \cite{gebert2015precision,solaro2020improved}, mostly NL effects arising only from the NP were considered in the King's plot to put bounds on $\alpha_{NP}$. For comparison, we first consider this procedure to infer bound on $\alpha_{NP}$ and refer to this as Case I analysis. In the next step, we account for all NL contributions given by Eq. (\ref{nonlinear_IS}) to infer constraint on $\alpha_{NP}$ and treat this as Case II procedure. We tabulate the inferred $\alpha_{NP}$ values from both the cases for a few selective $m_{\phi}$ values in Table \ref{NP_constraint} and compare them with the previously reported values for quantitative understanding. To weigh these bounds with other studies such as Casimir effects \cite{bordag2009advances}, $(g-2)$ factor of the electron \cite{fan2023measurement, morel2020determination} and neutron scattering \cite{nesvizhevsky2008neutron} and star cooling in a supernova \cite{raffelt2012limits}, we plot $\alpha_{NP}$ from all sources in Fig. \ref{bound}.

In Case I analysis, $\alpha_{NP}$ values were inferred at the 2$\sigma$ (95$\%$ confidence level) by combining our calculations with the experimental IS data for the $4S \rightarrow 3D_{5/2}$ and $4S \rightarrow 3D_{3/2}$ transitions, and using the procedure mentioned in Ref. \cite{solaro2020improved}. The bounds slightly differ from those reported in Ref. \cite{solaro2020improved, chang2024systematic}, where the uncertainty associated with $D$ was neglected. In this work, we account uncertainties to $D$ as the differences between both the AR and FF results. In Case II analysis, we account for the SM NL contributions, by subtracting the second-order FS and MS effects from the total IS data for the  $4S \rightarrow 3D_{5/2}$ transition obtained from \cite{solaro2020improved,chang2024systematic}. We obtain $\alpha_{NP} < 5.9 \times 10^{-10}$ for $m_{\phi} = 0$ eV. The contribution from $K^{(2)}$ is about 1.5 MHz, whereas that from $F^{(2)}$ is around 0.13 MHz—approximately an order of magnitude smaller. Here, we have assumed $\delta \langle r^2 \rangle^2\approx \delta\langle r^4\rangle$ owing to smaller size atomic nuclei of the Ca isotopes and used differential $\langle r^2 \rangle$ values from Ref. \cite{angeli2013table}.

We show bounds on $\alpha_{NP}$ for finite values of $m_{\phi}$ from both the AR and FF approaches from the present work and from other sources in Fig. \ref{bound}. As can be seen, the bound is almost independent of the boson mass for $m_{\phi}<10^3$ eV/c$^2$ where the interaction potential due to the boson goes inversely with the $e^- - n$ distance. This behavior is also evident from Table \ref{NP_constraint}, in which we have tabulated values explicitly. Further, for $m_{\phi}>10^5$ eV/c$^2$ the interaction range becomes shorter than the nuclear size causing the differences between the ratios $\frac{X_a}{F_a}$ and $\frac{X_b}{F_b}$ approach to zero, leading to a strongly suppressed sensitivity to new physics. In the intermediate mass range, accuracy in the calculated atomic wave functions becomes crucial and pronounced peaks in the bounds are seen between $m_{\phi}=10^3 $eV/$c^2$ and $10^4 $eV/$c^2$ in both Case I and Case II analyses. Bounds inferred from other sources are shown in the shaded regions of the above figure \cite{chang2024systematic, solaro2020improved, gebert2015precision, bordag2009advances, fan2023measurement, morel2020determination, nesvizhevsky2008neutron}, which show noticeable differences between atomic and other studies.

In summary, we have carried out rigorous investigations on the roles of higher-order non-linear contributions from the SM physics to the ISs of the singly ionized isotopes of Ca to reliably infer bounds on the electron-neutron coupling coefficient due to exchange of a vector boson. It demonstrates prominent roles of the second-order MS contributions for extracting the coefficient from the IS measurements accurately. It yields $\alpha_{NP} < 7.9 \times 10^{-11}$ when the higher-order SM non-linear effects were neglected and got elevated to $\alpha_{NP} < 5.9 \times 10^{-10}$ when these non-linear effects were accounted for in the analyses of the ISs in the Ca$^+$ ion.

BKS is supported by ANRF with grant no. CRG/2023/002558 and Department of Space, Government of India. All calculations reported in this work were performed on the ParamVikram-1000 HPC cluster at the PRL, Ahmedabad, Gujarat, India.

\bibliography{apssamp}

\end{document}